\documentclass[twocolumn]{aastex63}
\usepackage{amsmath}
\usepackage{color}
\usepackage{ulem}
\usepackage{graphicx}
\usepackage{textcomp}
\usepackage[T1]{fontenc}
\usepackage{footnote}
\usepackage{verbatim}

\begin{document}

\title{Nucleosynthesis contribution of neutrino-dominated accretion flows to the solar neighborhood}

\author[0000-0002-1768-0773]{Yan-Qing Qi}
\affiliation{Department of Astronomy, Xiamen University, Xiamen, Fujian 361005, China}
\affiliation{School of Physics and Astronomy, Monash University, Clayton, VIC 3800, Australia}
\affiliation{OzGrav: The ARC Centre of Excellence for Gravitational Wave Discovery, Clayton Victoria 3800, Australia}
\author[0000-0001-8678-6291]{Tong Liu}\thanks{E-mail: tongliu@xmu.edu.cn}
\affiliation{Department of Astronomy, Xiamen University, Xiamen, Fujian 361005, China}
\author[0000-0002-0771-2153]{Mouyuan Sun}
\affiliation{Department of Astronomy, Xiamen University, Xiamen, Fujian 361005, China}
\author[0000-0002-4223-2198]{Zhen-Yi Cai}
\affiliation{CAS Key Laboratory for Research in Galaxies and Cosmology, Department of Astronomy, University of Science and Technology of China, Hefei, Anhui 230026, China}
\affiliation{School of Astronomy and Space Science, University of Science and Technology of China, Hefei, Anhui 230026, China}

\begin{abstract}
The elemental abundances of stars reflect the complex enrichment history of the galaxy. To explore and explain the metal enrichment history of the cosmic environment near our solar system, we study the evolution of $^{56} \mathrm{Fe}$ abundance over time and [Mg/Fe] versus [Fe/H] evolution in the solar neighborhood. Core-collapse supernovae make the dominant contribution in the early stages, while Type Ia supernovae (SNe Ia) have a delayed and dominant impact in the later stages. In this work, we consider the nucleosynthesis contribution of neutrino-dominated accretion flows (NDAFs) formed at the end of the lives of massive stars. The results show that the [Fe/H] gradually increases over time and eventually reaches $\rm [Fe/H]=0$ and above, reproducing the chemical enrichment process in the solar neighborhood. Before the onset of SNe Ia, the ratio of $^{56} \mathrm{Fe}$ mass to the total gas mass increases by a factor of at most $\sim 1.44$ when NDAFs are taken into account. We find that by including NDAF in our models, the agreement with the observed metallicity distribution of metal-poor stars in the solar neighborhood ($\rm < 1~kpc$) is improved, while not significantly altering the location of the metallicity peak. This inclusion can also reproduce the observed evolutionary change of [Mg/Fe] at [Fe/H] $\sim -1.22$, bringing the ratio to match the solar abundance. Our results provide an extensive understanding of metallicity evolution in the solar environments by highlighting the nucleosynthesis contribution of NDAF outflows in the solar neighborhood.
\end{abstract}


\section{Introduction}

The chemical enrichment history of galaxies involves the complex formation and evolution of galaxies and stars and is one of the fundamental questions in astrophysics. Different elements are produced by stars and released into the galaxy on different timescales, so the abundance ratios of elements and isotopes within the galaxy change over time. The simplest galactic chemical evolution (GCE) model treats a galaxy as a one-zone chemical evolution model \citep[e.g.,][]{Schmidt1959,van1962,Schmidt1963,Tinsley1980,Timmes1995}, which assumes a closed system of primordial gas, a constant star initial mass function (IMF), chemically homogenous at any time, and instantaneous recycling \citep[e.g.,][]{Talbot1971,Searle1972}. However, the simple assumptions do not match up with all observational limits, like the metallicity distribution function (MDF) of the solar neighborhood (known as the G-dwarf problem with too many metal-rich stars relative to metal-poor stars in observations, see \citet{van1962} and \cite{Schmidt1963}). Multiple possibilities are proposed, including pre-enrichment \citep{Schmidt1963}, variable IMF or yields \citep{Truran1971}, inflow \citep[e.g.,][]{Larson1972,Tinsley1977,Chiappini1997,Kobayashi2000,Andrews2017}, various chemical enrichment sources \citep[e.g.,][]{Kobayashi2020a,Kobayashi2020b}, and multi-zone by representing the Milky Way as a series of concentric rings \citep[e.g.,][]{Chiappini1997,Portinari2000,Chiappini2001,Chen2023}.

An essential component of a GCE model is chemical yields synthesized in stellar evolution. The stellar yields are primarily affected by the mass and metallicity of the progenitors and initial explosion energy. Massive stars die quickly and release their yields and low-mass stars live a long time and retain most of their mass. Asymptotic giant branch (AGB) stars, with initial masses ranging from $0.5$ to $7~M_{\odot}$, produce C, N, O, and small amounts of s-process elements \citep[including about half the elements heavier than iron, e.g.,][]{Karakas2010,Kobayashi2011a}. Massive stars of 10 - 40 $M_{\odot}$ primarily produce $\alpha$ elements (O, Mg, Si, S, and Ca), which are ejected by core-collapse supernovae \citep[CCSNe, e.g.,][]{Timmes1995,Kobayashi2006}. Neutrino processes in CCSNe can enhance the production of certain elements, including F, K, Sc, and V \citep[e.g.,][]{Kobayashi2011b}. The Type Ia supernovae (SNe Ia), which are the explosions of carbon-oxygen (C+O) white dwarfs (WDs) in binary systems, produce half of the iron-peak elements (Cr, Mn, Fe, Ni, Co, Cu, and Zn). In the early stages of the galaxy, the $\rm [\alpha/Fe]$ ratio quickly reaches a plateau at about 0.5, mainly due to contributions from CCSNe. SNe Ia, which produce more Fe than $\alpha$-elements, start to appear around $\rm [Fe/H]\approx-1.0$ in the Milky Way. This delayed enrichment of SNe Ia causes $\rm [\alpha/Fe]$ to decrease as $\rm [Fe/H]$ increases \citep[e.g.,][]{Wallerstein1962,Tinsley1979,Matteucci1986,Kobayashi2009,Kobayashi2020b}. In addition, binary compact stars and accretion disk outflows are also considered as one of the sources of nucleosynthesis. The merger of binary neutron stars (NSs) or a black hole (BH) and an NS are thought to be the production sites of rapid neutron-capture process (r-process) elements, such as the lanthanides \citep[e.g.,][]{Li1998,Metzger2010,Metzger2019,Qi2022a,Kobayashi2023}. For the accretion disk outflows, \citet{Liu2021} discussed neutrino-dominated accretion flows (NDAFs) nucleosynthesis in the center of CCSNe, which may be a significant contributor to ${ }^{56} \mathrm{Ni}$ abundances for the progenitor stars with the mass range from 25 to 55 $M_{\odot}$. We found that the ratio of ${ }^{56} \mathrm{Fe}$ (produced by ${ }^{56} \mathrm{Ni}$ decays) mass to the initial total gas mass can be increased by a factor of no more than $1.95$ in terms of metal abundance evolution. \citet{Qi2022b} found that the quasar metallicity can reach about 20 times the solar metallicity for a top-heavy IMF if an upper limit on the nucleosynthesis yield of NDAF outflows and CCSNe are taken into account.

In recent years, a large amount of observational elemental abundance data in the Milky Way has been revealed by space astrometry missions (e.g., Gaia) \citep[e.g.,][]{Gaia2016,Gaia2018,Gaia2023}, medium-resolution multiobject spectroscopy surveys (e.g., the Apache Point Observatory Galactic Evolution Experiment (APOGEE) \citep[e.g.,][]{Majewski2017,Abdurro2022}, the High Efficiency and Resolution Multi-Element Spectrograph (HERMES) on the Anglo-Australian Telescope \citep{Raskin2011}, the 4 m Multi-Object Spectroscopic Telescope (4MOST) on the VISTA telescope \citep[e.g.,][]{de2012,de2019}, the William Herschel Telescope Enhanced Area Velocity Explorer \citep[WEAVE; e.g.,][]{Dalton2012,Jin2024}, and the Maunakea Spectroscopic Explorer \citep[MSE;][]{MSE2019}), and lower-resolution multiobject spectroscopy surveys (e.g., the Sloan Digital Sky Survey (SDSS) \citep[e.g.,][]{York2000,Abazajian2009}, the Radial Velocity Experiment (RAVE) \citep{Kunder2017}, the Large Sky Area Multi-Object Fibre Spectroscopic Telescope (LAMOST) \citep[e.g.,][]{Cui2012,Yao2019}, and the Prime Focus Spectrograph (PFS) on the Subaru Telescope \citep[e.g.,][]{Sugai2015,Tamura2016,Greene2022}). Stars form in clouds of gas enriched with heavy elements from previous stellar generations. As a result, the stellar atmospheres of non-evolved stars provide information about the metals present in the ISM during their formation \citep{Freeman2002}. These millions of stars provide direct observational evidence of the Milky Way's chemical evolution in space and time \citep[e.g.,][]{Casagrande2011,Bensby2014,Hayden2015,Zhao2016,Buder2018,Ahumada2020,Abdurro2022}.

To study the origin of metallicity, the solar neighborhood provides several significant advantages. First, the close distance of the Sun and other target stars offers high resolution and excellent observational data quality, which are crucial for accurately measuring the metallicity and other key physical parameters of stars. Second, the solar neighborhood contains various star types with different masses, ages, and evolutionary stages, providing ideal samples for systematically exploring the distribution of metallicity in different star types. In summary, the solar neighborhood is an ideal region to study metallicity and chemical evolution of galaxies due to its unique geographical and observational advantages, providing a foundation for the verification and calibration of GCE models, and improving the model's prediction of the behavior of other stellar regions.

In this paper, we build a solar neighborhood chemical evolution model by considering the additional nucleosynthesis contribution of NDAF outflows at the end of the lives of massive stars to gain a better understanding of ${ }^{56} \mathrm{Fe}$ and $\alpha$-elements (e.g., Mg) origins. In Section 2, we present the enrichment processes and describe the chemical evolution model. In Section 3, we show the results, which include ${ }^{56} \mathrm{Fe}$ evolution, MDF of ${ }^{56} \mathrm{Fe}$ and [Mg/Fe] versus [Fe/H] evolution in the solar neighborhood. A brief summary is made in Section 4.

\section{The Solar Neighbourhood Chemical Evolution Model}

\subsection{Chemical Enrichment Sources}

\begin{figure}
\centering
\includegraphics[width=1.0\linewidth]{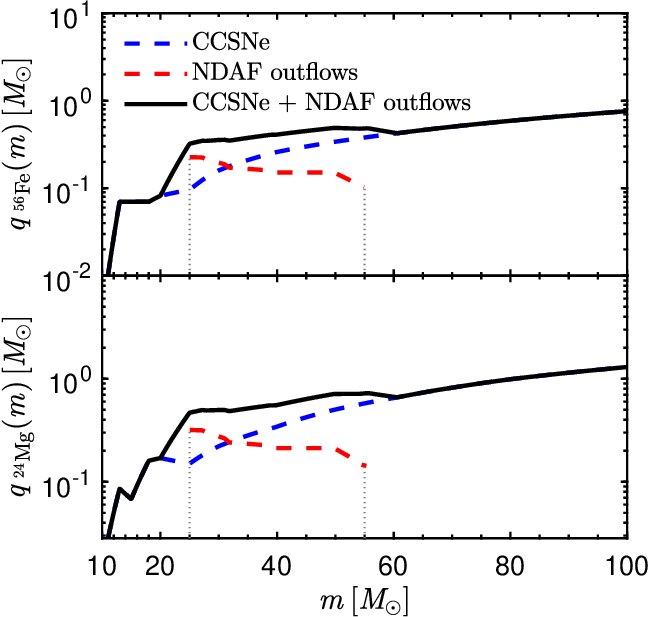}
\caption{${ }^{56} \mathrm{Fe}$ (top) and ${ }^{24} \mathrm{Mg}$ (bottom) yields from CCSNe and NDAF outflows as a function of progenitor-star mass. The blue dashed curve represents the yields from CCSNe, while the red dashed curve indicates the yields from NDAFs, whose progenitor-star mass ranges from 25 to 55 $M_\odot$. The black solid curve represents the total yields from the combined contributions of CCSNe and NDAF outflows. The CCSNe and NDAF outflows yields are adopted from \citet{Nomoto2013} and \citet{Liu2021}, respectively.}
\label{yields}
\end{figure}

We consider CCSNe, NDAF outflows, and SNe Ia as the metal processes to build a chemical evolution model, which aims to describe the ${ }^{56} \mathrm{Fe}$ abundance as a function of time in the solar neighborhood. The most significant uncertainty is the nucleosynthesis yields in each GCE model. Next, we summarize the key ingredient in the three chemical enrichment processes.

\emph{CCSNe} - Massive stars between $\sim$ 10 and $\sim 140~M_{\odot}$ will form an iron core and undergo core collapse. If this collapse is effective in causing an explosion, the stellar material will undergo shock heating and explosive nucleosynthesis. The mechanism for transitioning from collapse to explosion is not entirely understood \citep[e.g.,][]{Janka2012,Bruenn2013}. According to \citet{Nomoto2003}, stars with masses $\gtrsim$ 20 - 25 $M_{\odot}$ can form supernovae with two observational manifestations, one is the high-energy, bright hypernova branch, and the other is the low-energy, faint supernova branch. For faint supernovae, due to the substantially lower explosion energy $E_{51}=E/10^{51}~\mathrm{erg} < 1$, ${ }^{56} \mathrm{Ni}$ is yielded poorly and some of them might fall back into the compact remnant. In extreme cases, this branch becomes a failed supernova branch at larger progenitor star mass. \citet{Hamuy2003} pointed out that between these two major branches, there may be many types of supernovae. Furthermore, the problem of whether massive Type II supernovae (SNe II) can explode has been raised recently, both theoretically and observationally. Progenitor stars with initial masses $M>30~M_{\odot}$ have not been discovered in nearby SNe II-P \citep{Smartt2009}. It appears to be extremely difficult to explode stars larger than $25~M_{\odot}$ using the neutrino mechanism in multidimensional supernova explosion simulations \citep[e.g.,][]{Janka2012}; similar results are obtained with parameterized 1D models \citep{Ugliano2012,Muller2016}. Lower-metallicity stars may be more difficult to explode because their stellar cores are more compact than higher-metallicity stars. The fallback process can occur in spherical explosions with relatively low energy or in very energetic jet-like explosions \citep[e.g.,][]{Maeda2003,Tominaga2007,Tominaga2009}. In summary, stellar yields are influenced by various factors, such as the progenitor mass, metallicity, energy, and possible mixing and fallback. In this work, we use the ${ }^{56} \mathrm{Fe}$ and ${ }^{24} \mathrm{Mg}$ nucleosynthesis yields of CCSNe in \citet{Nomoto2013}, which provided a yields table (Yields Table 2013) of massive stars, listing the newly produced mass of each isotope up to ${ }^{74} \mathrm{Ge}$.

\emph{NDAF outflows} - In the center of a collapsar ($\gtrsim 25~M_\odot$), a stellar-mass BH with an accretion disk often forms. During the initial accretion phase, collapsar fallback matter triggers BH hyperaccretion processes that produce relativistic jets via the neutrino-antineutrino annihilation. This hyperaccretion disk is known as the neutrino-dominated accretion flow \citep[NDAF; for a recent review, see][]{Liu2017a}. As accretion rates decrease, the Blandford-Znajek (BZ) mechanism extracts rotational energy from the rotating BH and powers the Poynting flux \citep{Blandford1977}. In principle, the BZ mechanism could coexist and be the dominated mechanism of releasing energy in the whole life of the hyperaccretion process \citep[e.g.,][]{Lee2000,Globus2014,Pan2015}. The NDAF outflows contain abundant free protons and neutrons, which synthesize heavy metals in abundance. Between the relative proton-rich environments of collapsars and the neutron-rich conditions of compact object mergers, the products synthesized from NDAF outflows differ significantly \citep[e.g.,][]{Surman2006,Liu2013,Liu2021,Janiuk2014,Siegel2017}. The metallicity of the progenitor star or binary merger determines the electron fraction $Y_{e}$ at the outer boundary of the NDAF, thus affecting the type and amount of metals produced in the NDAF outflows \citep[e.g.,][]{Pruet2004,Surman2005,Surman2006,Liu2013,Liu2017b,Janiuk2014,Song2019}. In BH-NDAF systems at the centers of relative proton-rich collapsars, \citet{Liu2021} argued that approximately $10\%$ of the outflow material is synthesized into ${ }^{56} \mathrm{Ni}$ and gave the rough upper limits of the ${ }^{56} \mathrm{Ni}$ yields including the contributions of CCSNe and NDAFs to synthesize ${ }^{56} \mathrm{Ni}$. We found that by considering the NDAF outflow nucleosynthesis, the ${ }^{56} \mathrm{Ni}$ yields is weakly related to the progenitor mass. Moreover, the massive BH generated in the center of massive collapsars actually would reduce the efficiency of NDAFs until they cannot be ignited. Thus the mass of progenitors for BH-NDAF system have an upper limit. In this work, we also adopt the correlation between ${ }^{56} \mathrm{Fe}$ and ${ }^{24} \mathrm{Mg}$ yields per CCSN (i.e., the sum of the contributions of the conventional CCSNe and NDAFs outflows) and the progenitor star mass, ranging from 25 to 55 $M_{\odot}$, as shown in Figure~\ref{yields}.

\emph{SNe Ia} - The progenitor systems and the explosion mechanisms of SNe Ia remain a big debated problem \citep[e.g.,][]{Hillebrandt2000,Maoz2014,Soker2019}. The first scenario is known as the double-degenerate (DD) scenario, which involves the merging of two C+O WDs with a combined mass exceeding the Chandrasekhar mass limit \citep[e.g.,][]{Iben1984,Webbink1984}. The second scenario is called the single-degenerate (SD) scenario, which involves the accretion of hydrogen-rich matter through mass transfer from a binary companion until a single C+O WD reaches the Chandrasekhar mass limit and explodes \citep[][]{Whelan1973}. The binary systems of C+O WDs with main-sequence (MS) or red-giant (RG) secondary stars are adopted as progenitor systems. MS+WD systems dominate star-forming galaxies with timescales of $\sim$ 0.1 - 1 Gyr, while RG+WD systems have lifetimes of $\sim$ 1 - 20 Gyr and dominate among early-type galaxies. For the nucleosynthesis yields of SNe Ia, we adopt the Chandrasekhar mass model W7 \citep{Nomoto1997}, which is satisfied with SNe Ia observed spectra and GCE in the solar neighborhood \citep{Kobayashi2006}.

\subsection{Model}

We consider a chemical evolution model of the solar neighborhood \citep[e.g.,][]{Tinsley1980,Liu2021}, initializing the $\rm 1~kpc$ solar vicinity with $M_{0, \text { cold }}=1.0 \times 10^{8} M_{\odot}$ of cold gas. Star formation (SF) rate is often calculated using the Kennicutt-Schmidt law \citep[$\dot{\Sigma}_{\star} \propto \Sigma_{\mathrm{gas}}^{N}$,][]{Schmidt1959,Kennicutt1998}. An extensive study of SF rates in the global galaxy by \citet{Kennicutt1998} found that the best-fitting power law index of $N=1.4$, implying a higher SF efficiency in denser regions. \citet{Leroy2008} and \citet{Bigiel2008} found a value of $N$ of 1.0 for molecularly-dominated gas from spatially-resolved observations of local galaxies, suggesting that sufficiently dense gas has a constant SF efficiency. \citet{Scoville2016} measured the total gas mass using the dust continuum and found a linear slope ($N=0.9$) for star-forming galaxies at higher redshifts ($z \sim 1{-}6$). Our model assumes a constant effective SF rate ($N=1.0$):
\begin{equation}
\dot \Psi = M_{\rm g} ~ / ~ t_\star,
\end{equation}
where $t_\star=4.70~\rm Gyr$ is a typical SF timescale, $ M_{\rm g}$ is the total gas mass of the solar neighbourhood.

The following equations then describe the time variations of gas mass $M_{\rm g}$ and abundance of element $i$:
\begin{equation}
\frac{d M_{\rm g}}{dt} = - \dot \Psi + R + R_{\rm Ia} + R_{\rm in},
\end{equation}
and
\begin{equation}
\frac{d(M_{\rm g} X_{i})}{dt} = - \dot \Psi X_{i} + E_{i} + E_{i, \rm Ia} + R_{\rm in}X_{i,\rm in},
\end{equation}
where $X_i$ is the mass fraction of element $i$. The metallicity of the infalling gas $X_{i,\rm in}$ is assumed to be 0.

The infall rate $R_{\rm in}$ with an infall timescale $\tau_{\mathrm{i}}=5.0~\rm Gyr$ of primordial gas from outside the system is
\begin{equation}
R_{\rm in}= \frac{t}{\tau_{\mathrm{i}}} \exp \left[-\frac{t}{\tau_{\mathrm{i}}}\right]
\end{equation}
for the solar neighborhood \citep{Pagel2009}.

The gas ejection rate from dying stars at time $t$ is
\begin{equation}
R(t) = \int_{m_{\star}(t)}^{m^{\rm max}} (m - m_{\rm rem})~ \phi(m) ~ \dot \Psi[t - \tau(m)] dm,
\end{equation}
where $m_{\rm rem}$ is the dimensionless remnant mass, and $m_\star$ is the dimensionles mass of the star for which $\tau(m_\star) = t$. We adopt the IMF from \citet{Kroupa2001}, which is a power-law mass spectrum $\phi(m)\equiv dN/dm\propto m^{-\alpha_{i}}$ with $\alpha_{1}=1.3$ for $m^{\rm min}=0.08\leq m \leq0.5$, and $\alpha_{2}=2.3$ for $0.5\leq m \leq100=m^{\rm max}$. It is normalized as $\int_{m^{\rm min}}^{m^{\rm max}} m\phi(m) dm=1$. The remnant mass is given by \citep[e.g.,][]{Weidemann1983,Iben1984,Thorsett1999,Pagel2009}
\begin{equation}
m_{\mathrm{rem}} \approx\left\{\begin{array}{ll}
0.106 m+0.446, & 0.5<m<9 \\
1.4, & 9<m<25 \\
0.24 m-4. & m>25
\end{array}\right.
\end{equation}

The Geneva group \citep{Schaller1992} estimates the lifetime of a star with mass $m$ as
\begin{equation}
\tau(m) \approx 11.3 m^{-3} + 0.06 m^{-0.75} + 0.0012~~{\rm Gyr}.
\end{equation}

The total ejection rate of element $i$ is obtained by
\begin{equation}
\begin{aligned}
E_{i}(t) \simeq & \int_{m_{\star}(t)}^{m^{\max }}\left\{\left(m-m_{\mathrm{rem}}\right) X_{i}[t-\tau(m)]+q_{i}(m)\right\} \\
& \times \phi(m) \dot \Psi[t-\tau(m)] d m,
\end{aligned}
\end{equation}
which includes both returned fraction (unprocessed metals) to the interstellar medium (ISM) and newly synthesized elements $q_{i}(m)$ \citep{Nomoto2013} as shown in Figure~\ref{yields}, we specifically consider ${ }^{56} \mathrm{Fe}$ (produced by the decay of ${ }^{56} \mathrm{Ni}$) and ${ }^{24} \mathrm{Mg}$. And the solar abundances are taken from \citet{Asplund2009}.

$N_{\rm{Ia}}(t)$ is the total number of SNe Ia per year \citep[e.g.,][]{Kawata2003,Few2014,Snaith2015}:
\begin{equation}
\begin{aligned}
N_{\rm{Ia}}(t)= & \int_{\max \left[m_{\rm{p}, \ell}, m_{\star}(t)\right]}^{m_{\rm{p}, u}} \phi(m) \dot \Psi[t-\tau(m)] dm \\
& \times\left[ b_{\rm{MS}} \frac{\int_{\max\left[m_{\rm{MS}, \ell}, m_{\star}(t)\right]}^{m_{\rm{MS}, u}} \phi(m) \, dm}{\int_{m_{\rm{MS}, \ell}}^{m_{\rm{MS}, u}} \phi(m) \, dm} \right.\\
& \left.+b_{\rm{RG}} \frac{\int_{\max \left[m_{\rm{RG},\ell},  m_{\star}(t)\right]}^{ m_{\rm{RG},u}} \phi(m) dm}{\int_{m_{\rm{RG}, \ell}}^{m_{\rm{RG},u}} \phi(m) dm}\right].
\end{aligned}
\end{equation}

\begin{figure}
\centering
\includegraphics[width=1.0\linewidth]{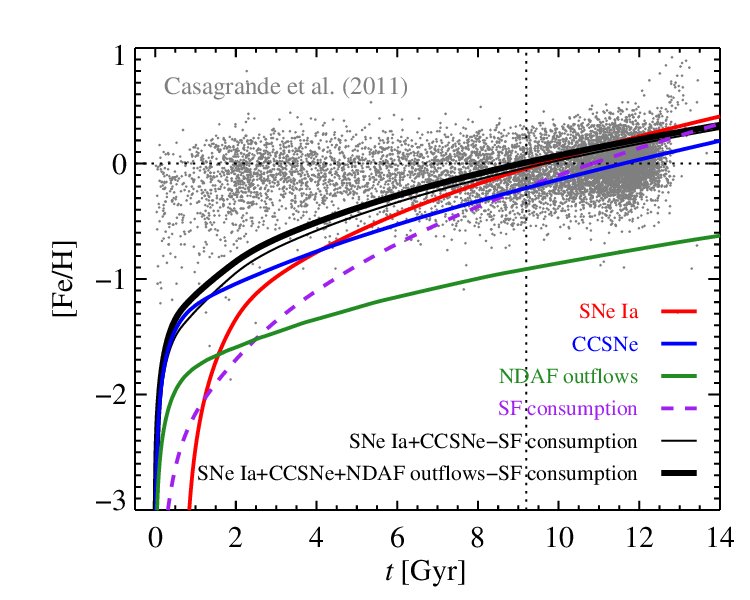}
\caption{Metallicity evolution of ${ }^{56} \mathrm{Fe}$ in the solar neighborhood, shown for two independent cases: one with CCSNe combined with NDAF outflows (black thick curve) and another without NDAF outflows (black thin curve). Colorful curves represent the contribution of different nucleosynthetic processes: SNe Ia (red solid curve), CCSNe (blue solid curve), and NDAF outflows (green solid curve). The purple dashed curve illustrates the consumption effect of SF processes on metallicity. The grey dots represent observational thin disk stars within 1 kpc from \citet{Casagrande2011}. The grey dashed line represents the solar metallicity at the time of the Sun's birth. The Kroupa IMF parameters are used with $m^{\rm min} = 0.08~M_{\odot}$, $m^{\rm max} = 100~M_{\odot}$, and $t_{\star} = 4.70~\rm Gyr$.}
\label{Metallicity evolution}
\end{figure}

The integrals are computed independently for the primary and secondary stars. The primary stars are assumed to have an MS mass range between $m_{\rm {p},\ell}=3~M_{\odot}$ and $m_{\rm {p},u}=8~M_{\odot}$ and form C+O WDs. Secondaries are either MS with mass range from $m_{\rm{MS}, \ell}=1.8~M_{\odot}$ to $m_{\rm{MS}, u}=2.6~M_{\odot}$ or RG with mass range from $m_{\rm{RG}, \ell}=0.9~M_{\odot}$ to $m_{\rm{RG}, u}=1.5~M_{\odot}$. The lifetimes of SNe Ia, representing the lifetimes of binary systems from their formation to the explosion, are determined by the lifetimes of secondary stars. But \citet{Kobayashi2009} pointed out that the $\rm [Fe/H]$ at which $\rm [\alpha/Fe]$ starts to decrease depends mainly on the metallicity dependence of SNe Ia progenitor rather than lifetime. The binary fraction for each of the secondary types is $b_{\rm MS} = 0.04$ and $b_{\rm RG}=0.02$ \citep{Kobayashi2020a}.

The gas ejection rate of SNe Ia is
\begin{equation}
R_{\rm Ia}(t)=m_{\rm{CO}} N_{\rm{Ia}}(t),
\end{equation}
where the mass of the WD at the explosion is denoted by $m_{\rm CO}=1.38~M_\odot$.

Element $i$ is ejected by SNe Ia at a rate of $E_{i, \rm Ia}$ as follows
\begin{equation}
E_{i, \rm Ia}(t)=q_{i, \mathrm{Ia}}N_{\rm{Ia}}(t),
\end{equation}
where the ${ }^{56} \mathrm{Fe}$ and ${ }^{24} \mathrm{Mg}$ yields are approximately $0.613~M_{\odot}$ and $0.0085~M_{\odot}$, respectively \citep{Nomoto1997}.

\subsection{MDF}

We consider the distribution of [Fe/H] as a proxy of MDF. Given the SF rate $\dot \Psi(t)$, the amount of stellar mass $dM$ formed in the time interval $[t, t + dt]$ is
\begin{equation}
dM=\dot \Psi (t)dt.
\end{equation}

By definition of normalized IMF $\phi(m)$, for a population of mass $dM$, the total number of present-day stars with masses $m^{\rm min}\leq m \leq m^{\star,\rm max}$ is
\begin{equation}
dN_{\star}=dM\int_{m^{\rm min}}^{m^{\star,\rm max}(t)} \phi(m) dm,
\end{equation}
where the integration limits includes all the possible stellar masses that were born at time $t$ and can survive to the present day, excluding sources more massive than $m^{\star,\rm max}$.

The number of present-day stars in the $\mathit{i}$th metallicity bin $\rm [Fe/H]_{\mathit{i}}\leq[Fe/H](\mathit{t})\leq[Fe/H]_{\mathit{i}+1}$ is
\begin{equation}
\Delta N_{\star}=\int_{t_{i}}^{t_{i+1}}dN_{\star},
\end{equation}
which provides tighter observational constraints on the evolution of metallicity in the solar neighborhood.

\section{Results}

\begin{figure}
\centering
\includegraphics[width=0.95\linewidth]{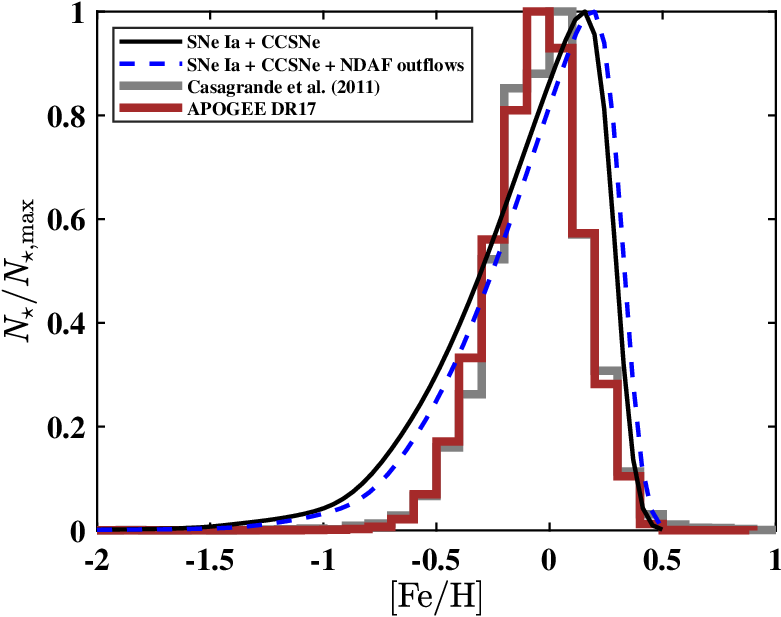}
\caption{The present-day MDFs in cases of CCSNe combined with (black solid curve) or without (blue dashed curve) NDAF outflows in the solar neighborhood. The grey empty histogram and brown empty histogram represent the MDF of observed solar neighborhood stars from thin disk (within the 1 kpc) taken from \citet{Casagrande2011} and APOGEE DR17 \citep{Abdurro2022}, respectively.}
\label{MDF}
\end{figure}

Figure~\ref{Metallicity evolution} shows the evolution of [Fe/H] as a function of time in the solar neighborhood. The various astrophysical sources' contributions to the evolution of ${ }^{56} \mathrm{Fe}$ are shown as a red solid curve (SNe Ia), blue solid curve (CCSNe; including the unprocessed metals fraction returned into the ISM, and new elements produced by nucleosynthesis), green solid curve (NDAF outflows), and purple dashed curve (SF consumption). The total evolution of ${}^{56}\mathrm{Fe}$ is shown by the black thick curve when NDAF outflows are included, and by the black thin curve when NDAF outflows are excluded. The red, blue, and purple components contribute to both the black thick and the black thin curves, while the green NDAF component contributes only to the black thick curve. It is noteworthy that the trend of [Fe/H] in the initial time is a rapid increase, then followed by a slow increase, reflecting the different effects of stellar activity on the metal enrichment of ISM. In the early stage, CCSNe and NDAF outflows mainly drive the increase in metallicity. The contribution of NDAF outflows is comparable to the fresh yields of CCSNe, indicating that both NDAF outflows and CCSNe nucleosynthesis play an important role in the overall metallicity evolution. Since NDAFs may be formed in the center of CCSNe, the environment favorable to CCSNe therein may also promote NDAFs. Of course, the low initial explosion energy of CCSNe could enhance the inflow and outflow processes, but is to the disadvantage of the metal diffusion. At $\rm 4.11~Gyr$, the role of SNe Ia becomes prominent and dominates the metal enrichment in the later stage. This transition highlights the delayed impact of SNe Ia on the chemical evolution of galaxies. At approximately 0.01 Gyr, the [Fe/H] abundance without NDAFs is $-3.20$, whereas with NDAFs, it increases to $-3.04$. This indicates that NDAFs can cause an increase in the ratio of ${ }^{56} \mathrm{Fe}$ mass to the total gas mass by a maximum factor of $\simeq 1.44$. The presence of NDAFs does not alter the contributions from other sources (SNe Ia, CCSNe, SF consumption) but adds an additional source of iron enrichment, leading to a higher [Fe/H] ratio in the early stages.

\begin{figure}
\centering
\includegraphics[width=1.0\linewidth]{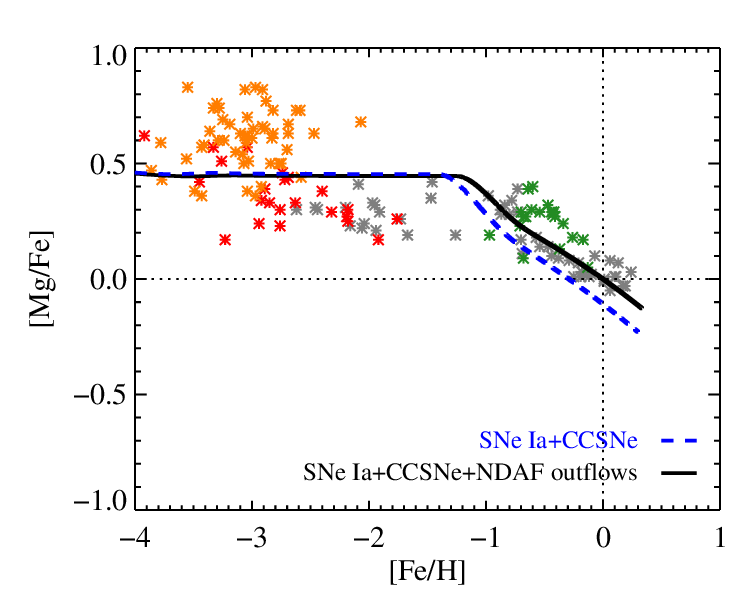}
\caption{[Mg/Fe] versus [Fe/H] evolution for the solar neighborhood with (black solid curve) or without (blue dashed curve) NDAF outflow contributions. The observational data are obtained with NLTE analysis from \citet{Andrievsky2010} (orange asterisks), \citet{Mashonkina2017,Mashonkina2019} (red and green asterisks), and \citet{Zhao2016} (grey asterisks).}
\label{[Mg/Fe]}
\end{figure}

The metallicity with or without NDAF outflows contribution reaches [Fe/H] = 0 at $\rm9.08~Gyr$ and $\rm9.66~Gyr$, respectively. The inclusion of NDAF outflows demonstrates that [Fe/H] reaches the solar ratio ($\rm [Fe/H] = 0$) at the time of the Sun's formation (4.70 Gyr ago), indicating that the metallicity of the average ISM near the Sun at that time is comparable to the Sun. The grey dots are observational data from \citet{Casagrande2011} in the solar neighborhood. From the perspective of additional sources of metals, our model provides a new explanation for a large number of young stars that exhibit super-solar metallicity in observations. These results emphasize the non-negligible role of NDAF outflows in the metal enrichment process of the ISM. While our model in the [Fe/H] - $t$ diagram can effectively explain the metallicity of younger stars, it encounters difficulties in accounting for the high metallicity observed in older stars. Theoretical GCE models are typically based on the assumption of homogeneous chemical evolution within galaxies. However, a range of factors can also influence the actual metallicity of stars, including initial galactic conditions, local non-homogeneous mixing of star clusters, the history of galaxy mergers and evolution, as well as observational limitations and selection effects. These factors make it challenging to use the model's [Fe/H] evolution at early times to explain the presence of metal-rich old stars in observations.

Figure~\ref{MDF} shows the MDF of present-day stars in the solar neighborhood with (black solid curve) and without (blue dashed curve) the nucleosynthesis contribution from NDAF outflows. Based on the three Galactic space velocities, U (the velocity towards the Galactic centre), V (the velocity along the Galactic rotation), and W (the velocity in the direction of the North Galactic Pole) \citep{Li2017}, stars belonging to the thin disk can be identified. The grey and brown empty histograms represent the observational data of thin disk stars within the solar neighborhood ($<1~\rm kpc$) from \citet{Casagrande2011} and APOGEE DR17 \citep{Abdurro2022}, respectively. We find that including NDAF outflows in our models shifts the MDF slightly toward the more metal-rich side, while the peak remains close to solar metallicity ($\rm [Fe/H] = 0$). The modeled MDF predicts a greater number of both metal-poor and metal-rich stars than observed. Additionally, our model exhibits a skewed-negative MDF, similar to the weakly skewed-negative form seen in the observational data. Overall, the modeled MDF roughly reproduces the peak [Fe/H] and the general shape of the observed MDF, improving the agreement with the observed metallicity distribution of metal-poor stars when NDAF outflows are included, while not significantly altering the location of the metallicity peak.

Figure~\ref{[Mg/Fe]} illustrates the evolution of [Mg/Fe] versus [Fe/H] in the solar neighborhood, comparing scenarios with (black solid curve) and without (blue dashed curve) contribution of NDAF outflows. Observational data are represented by differently colored asterisks: orange \citep{Andrievsky2010}, red and green \citep{Mashonkina2017,Mashonkina2019}, and grey \citep{Zhao2016}. In the early stages of galaxy formation, the [Mg/Fe] ratio forms a plateau across a wide range of [Fe/H]. With NDAF outflows, this plateau reaches a value of approximately 0.44, slightly lower than the 0.45 plateau in the case without NDAF outflows. The plateau extends up to [Fe/H] $\approx -1.22$ with NDAF outflows and up to [Fe/H] $\approx -1.37$ without NDAF outflows, after which the [Mg/Fe] ratio decreases sharply. The [Fe/H] value at which this decline begins depends on the adopted SNe Ia progenitor model, specifically the progenitor lifetime. Around [Fe/H] $\approx -1.0$, SNe Ia begin contributing significantly, producing more iron relative to $\alpha$-elements, which causes a decrease in the [Mg/Fe] ratio as [Fe/H] increases. When NDAF outflows are included, our model aligns more closely with observational trends, eventually approaching solar ratios ($\rm [Mg/Fe] = [Fe/H] = 0$). While the baseline GCE model without NDAF outflows gives an [Mg/Fe] value of $-0.10$ when $\rm [Fe/H] = 0$. \citet{Palla2022} predicted [Mg/Fe] values above $-0.20$ at solar iron abundance are generally favored. Including NDAF outflows in our model roughly reproduces the slope of [Mg/Fe] and aligns with solar ratios.

\section{Summary}

Our models characterize the nucleosynthesis contributions of NDAF outflows of the ${ }^{56} \mathrm{Fe}$ and [Mg/Fe] vs. [Fe/H] evolution in the solar neighborhood. The evolution of metallicity shows an increasing trend over time and reaches $\rm [Fe/H]=0$ at the time of $\sim \rm 9.08~Gyr$, dominated by CCSNe and NDAF outflows in the early stages and by the delayed SNe Ia in the later stages from $\sim \rm4.11~Gyr$. At $\sim 0.01$ Gyr, the ratio of ${ }^{56} \mathrm{Fe}$ mass to the total gas mass increases by a maximum of $\sim 1.44$ times if NDAFs are considered. It is noteworthy that the NDAF is theorized to occur in the center of some CCSNe. And the NDAF and CCSNe may have a common and important role in promoting a metal-rich ISM. Including NDAF outflows in the modeled MDF of present-day stars enhances the agreement with the observed metallicity distribution of metal-poor stars in the solar neighborhood ($\rm < 1~kpc$), while still reproducing overall shape of the observed MDF without significantly altering the metallicity peak's location. The results indicate that incorporating NDAF outflows in a one-zone GCE model calibrated to fit the Galactic properties improves the agreement with iron abundance distributions observed in a large sample of solar neighborhood stars. Moreover, including NDAF outflows in our models results in a [Mg/Fe] ratio that more closely aligns with observational data by extending the plateau to 0.44 up to [Fe/H] $\sim$ $-1.22$ and enabling the ratio to eventually reach solar abundance values ($\rm [Mg/Fe] = [Fe/H] = 0$). Our GCE model considering NDAF outflows as an additional metal source roughly reproduces various observational constraints on the chemical evolution of the solar neighborhood, such as the [Fe/H] ratio trends, the MDFs, and [Mg/Fe] vs. [Fe/H] evolution. While the results of the GCE model still exhibit inconsistencies with observations in explaining the high metallicity of older stars. In the future, James Webb Space Telescope (JWST) will release a large number of high-redshift quasars, whose metallicity can better reflect the metal enrichment process in the early universe and provide observational constraints for the galaxy formation and evolution. We will construct a GCE model of early galaxies and constrain their IMFs to further explore the impact of NDAF outflow nucleosynthesis on these early galaxies.

\acknowledgments
We thank the anonymous referee for very helpful suggestions and comments and Ling-Lin Zheng and Jun-Hui Liu for helpful discussions. This work was supported by the National Key R\&D Program of China under grants 2023YFA1607902 and 2023YFA1607903), the National Natural Science Foundation of China under grants 12173031, 12221003, 12322303, and 12373016, the Natural Science Foundation of Fujian Province of China (No. 2022J06002), and the Fundamental Research Funds for the Central Universities (No. 20720240152).

{}

\end{document}